\documentclass[aps,twocolumn]{revtex4-1}
\usepackage{latexsym}
\usepackage{epsfig,pstricks,graphicx}
\usepackage{amsmath,amssymb}
\usepackage[mathscr]{eucal}
\usepackage{color}

\def\DJo{$\;$\kern-.4em \hbox{D\kern-.8em\raise.15ex\hbox{--}\kern.35em okovi\'c}}
\def\CC{{\rm\kern.24em \vrule width.04em height1.46ex depth-.07ex
\kern-.30em C}}
\def\RR{{\rm
         \vrule width.04em height1.58ex depth-.0ex
         \kern-.04em R}}
\def\P{{\rm I\kern-.25em P}}
\def\id{{\rm 1\kern-.22em l}}
\usepackage{amsmath}
\usepackage{amssymb}
\usepackage[T1]{fontenc}
\usepackage{graphicx}

\newcommand{\bra}[1]{\ensuremath{\left<#1\right|}}
\newcommand{\ket}[1]{\ensuremath{\left|#1\right>}}

\newcommand{\rme}{\ensuremath{\mathrm{e}}}
\newcommand{\rmi}{\ensuremath{\mathrm{i}}}

\newcommand{\tr}{\operatorname{tr}}
\newcommand{\diagSL}[1]{\left(\begin{matrix}
      #1&0\\0&1/#1
    \end{matrix}\right)}
\newcommand{\diag}{\operatorname{diag}}

\begin{document}
\title{Rescaling multipartite entanglement measures for mixed states
      }
\author{Oliver Viehmann,$^1$ 
        Christopher Eltschka,$^2$ 
        and Jens Siewert$^{3,4}$\vspace*{0.1cm}
}                   

\affiliation{$^1$Physics Department,
             Arnold Sommerfeld Center for Theoretical Physics,
             and Center for NanoScience,\\
             Ludwig-Maximilians-Universit\"at,
             Theresienstra{\ss}e 37,
             D-80333 M\"unchen, Germany}
%
%
\affiliation{$^2$Institut f\"ur Theoretische Physik, 
             Universit\"at Regensburg, D-93040 Regensburg, Germany}
%
%
\affiliation{$^3$Departamento de Qu\'{\i}mica F\'{\i}sica, 
             Universidad del Pa\'{\i}s Vasco, 
             Apdo.\ 644, 48080 Bilbao, Spain}
%
%
\affiliation{$^4$IKERBASQUE, Basque Foundation for Science, 
            Alameda Urquijo 36, 48011 Bilbao, Spain}

%
%

\begin{abstract}
A relevant problem regarding entanglement measures is the following: 
Given an arbitrary mixed state,
how does a measure for multipartite entanglement change if general local operations are
applied to the state? This question is nontrivial as the normalization of the states
has to be taken into account. Here we answer it for pure-state
entanglement measures which 
are invariant under determinant-one local operations and homogeneous in the state coefficients,
and their convex-roof extension which quantifies mixed-state entanglement.
Our analysis allows us to enlarge the set of mixed states for which these important measures 
can be calculated exactly.
In particular, our results hint at a distinguished role of entanglement measures
which have homogeneous degree 2 in the state coefficients.
\end{abstract}

\maketitle

\section{Introduction}
In recent years, there has been astounding experimental progress in preparing and operating
multiqubit entanglement, e.g., with trapped ions~\cite{Blatt14ions2011,FSK2005}, 
photons~\cite{Pan2007,Weinfurter2009}, and superconducting circuits~\cite{Martinis2010,Schoelkopf2010}.
On the other hand, entanglement theory is lagging behind this advancement. To date, it is not
possible to adequately characterize the entanglement of an experimentally prepared multipartite quantum state, 
that is, to quantify how much of a certain type of entanglement is contained in that state.

One difficulty here is that it is not enough to just decide ``how much entanglement is present''.
Entanglement is regarded as a resource for quantum information tasks, and different tasks
require different resources, that is, inequivalent types of entanglement. The criterion whether
or not two states can be considered equivalent with respect to their entanglement
is whether they can be transformed into one another by means of stochastic local
operations and classical communication (SLOCC)~\cite{Bennett2000,DVC2000}. 
The consequence is that different equivalence classes of entanglement need 
to be distinguished~\cite{DVC2000,Verstraete2002,Lamata2006,Bastin2009}.

Entanglement witnesses provide a tool to distinguish SLOCC classes of
multipartite states and to experimentally detect
them~\cite{GuehneToth2009}. There are attempts to make witnesses more
quantitative~\cite{Reimpell2007,Eisert2007} which 
have not found widespread application in practice yet.

So far there are only few practically relevant tools for measuring the entanglement of general (mixed) quantum states.
Negativity-related entanglement monotones~\cite{Vidal2002,Plenio2005} represent one of them. 
They are applied only to bipartite systems and it is not clear if they can be used to distinguish 
different types of entanglement. Therefore
we do not further investigate them here. 
The second well-established tool is Wootters' approach to compute mixed state entanglement for 
two qubits~\cite{Wootters1998}.
It is based on the concurrence, a pure-state entanglement measure which
is generalizable to pure states of systems larger than two qubits. The 
kind of pure-state entanglement measures the concurrence belongs to has several 
interesting properties (see below). In particular, 
entanglement measures of this kind can distinguish
and quantify specific types of multipartite entanglement.
Re\-cent\-ly there has been progress regarding the question how to obtain
SLOCC classifications using them~\cite{Shelly2010,Viehmann2011,LiLi2011}.
An application of these quantities to mixed states in analogy to Wootters' method is therefore
highly desirable.

The concurrence is
an example of a {\em polynomial invariant}, i.e., a polynomial function of the 
coefficients (and {\em not} their complex conjugates) of a quantum state, where
the function is invariant under determinant-one local operations,
that is, under the transformations 
\begin{eqnarray}
\label{genlocop}
A & = & A_1\otimes \ldots \otimes A_N\nonumber\\
  & \equiv & \dfrac{F_1}{(\det F_1)^{{1}/{d_1}}} \otimes \cdots  \otimes \dfrac{F_N}{ (\det F_N)^{{1}/{d_N}}} \ \ .
\end{eqnarray}
Here the operators $F_k \in \mathrm{GL}(d_k,\mathbb{C})^{\otimes N}$ represent 
general (invertible) local  operations 
on the $N$-qudit Hilbert space $\mathcal{H} = \bigotimes_{k=1}^N \mathbb{C}^{d_k}$.
Obviously,  $A_k$ have determinants equal to 1, i.e., $A_k\in \mathrm{SL}(d_k,\mathbb{C})$.

The peculiarity of polynomial invariants is that a power of them which has homogeneous
degree 2 in the state coefficients 
is  automatically an entanglement monotone~\cite{Verstraete2003}.
The generalizations of the concurrence, that is,
polynomial invariants for pure states of three and more qubits are 
well known~\cite{CKW2000,Wong2001,LuqueThibon2003,OS2005,Dokovic2009}.

Hence, pure-state entanglement measures are known in principle 
at least for multiqubit systems.
It is straightforward to generalize them to mixed states via the convex-roof 
extension~\cite{Uhlmann1998} (for details, see below). The problem of quantitative
entanglement theory is that it is not known how to compute the convex roof of an
invariant in general
(or even a nontrivial lower bound of it) for any multipartite system, except 
for the concurrence of two qubits and its straightforward generalization
to an even number of qubits, the {\em N-concurrence}~\cite{Uhlmann2000,Bullock2004}.

At least there exist a few exact solutions for three-qubit states of low rank and high 
symmetry~\cite{Lohmayer2006,Eltschka2008,Jung2009,Fei2011}.
To study how to extend these solutions to a larger family of states 
is part of the quest for a better understanding of the convex roof and 
multipartite entanglement in mixed states.
An important question in this context is
how an entanglement measure behaves under general local operations. While 
local unitary operations represent mere basis changes and do
not change entanglement, general local operations require
subsequent normalization of the state and thus a conversion of the amount of 
entanglement. 

In this work we investigate how multipartite entanglement
needs to be rescaled if it is quantified by functions which are
homogeneous and invariant under determinant-one SLOCC operations. 
It will turn out that in the case of mixed states this is easily done only for 
the convex roof of pure-state measures with homogeneous 
degree 2. 
That is, from the point of view of mixed-state entanglement
rescaling, one can conclude that there is a preferred homogeneity degree of
pure-state entanglement measures. 

The outline of this article is as follows.
First we analyze the rescaling properties of the
convex-roof extension. Then
we illustrate their application (and also their failure for the ``wrong''
homogeneity degree) by reconsidering some of the known exact solutions. Finally, we
mention also some ways to extend the number of exactly solvable three-qubit
problems.

\section{The rescaling method}\label{sec:rescalingmethod}
\label{sect:rescalingmethod}

This section provides the theoretical concepts on which this article is based. They are formulated for general homogeneous invariants, but we interpret them with regard to the physically relevant polynomial $\mathrm{SL}(d_k, \mathbb{C})^{\otimes N}$-invariants. To begin with, we introduce our notation and explain how a homogeneous invariant of pure states needs to be rescaled under a symmetry transformation if one exclusively focuses on normalized states. Then we explain the convex-roof extension. Finally, we state, derive, and discuss our main results. They describe the scaling of the convex-roof extension of a homogeneous invariant under a symmetry transformation for normalized mixed states.

\subsection{Rescaling of homogeneous invariants for normalized pure states}
We consider a real function $\tau$ on some finite-dimensional Hilbert space $\mathcal{H}$ that is invariant with respect to some invertible linear operator $L$ acting on $\mathcal{H}$. In formulas, $\tau (L \psi)  = \tau (\psi) $ for all $\psi \in \mathcal{H}$. This implies $\tau(L^{-1} \psi) = \tau(\psi)$. We assume further that $\tau$ is homogeneous of degree $\eta$. That is, for positive $\alpha$, $\tau(\alpha \psi) = \alpha^\eta \tau(\psi)$. If we focus on normalized states $\phi$ and $\phi^\prime$ which are related by $L$ according to $\phi^\prime = L \phi/ || L \phi ||$, it follows that
\begin{align}\label{eq:result_pure_states}
\tau(\phi^\prime) = \dfrac{\tau(\phi)}{||L \phi ||^\eta}.
\end{align}
Note that $\tau$ could be one of the polynomial invariants discussed in the introduction
(or a function thereof) if $\mathcal{H} = \bigotimes_{k=1}^N \mathbb{C}^{d_k}$ and 
$L\equiv  A \in \mathrm{SL}(d_k,\mathbb{C})^{\otimes N}$. 
Note further that Eq.~\eqref{eq:result_pure_states} 
has the same form for all $\eta$, so that there is no preferred homogeneity degree for the polynomial invariant.

\subsection{The convex roof extension}
Let $\mathcal{H}$ be a finite-dimensional Hilbert space and $\Omega_\mathcal{H}$ the convex set of (normalized) density operators acting on $\mathcal{H}$. All $\rho \in \Omega_\mathcal{H}$ can be written as convex sums $\rho = \sum_i p_i \pi_i$.  That is, $p_i \geq 0$, $\sum_i p_i = 1$, and $\pi_i$ is an extreme point of $\Omega_\mathcal{H}$. An extreme point $\pi$ of $\Omega_\mathcal{H}$ is given by $\pi = \ket{\psi} \! \bra{\psi}/||\psi ||^2$, where $\psi \in \mathcal{H}$ ($\neq 0$), and is usually referred to as a pure state. A mixed state $\rho$ has rank $> 1$, and its decomposition $\{ p_i, \pi_i \}$ into pure states is not unique \cite{Schroedinger1936}.

A real continuous function $f$ on $\mathcal{H}$ can be extended to $\Omega_\mathcal{H}$ via \textit{convex-roof extension}~\cite{Uhlmann1998}:
\begin{align}\label{eq:definition_convex_roof}
f (\rho) = \mathrm{min} \sum_i p_i f (\pi_i).
\end{align}
The minimum runs over all possible decompositions of $\rho$ into pure
states, and $f(\pi_i) \equiv f (\psi_i/ ||\psi_i||)$. A decomposition $\{
\tilde{p}_i, \tilde{\pi}_i \}$ of $\rho$ for which $f(\rho) = \sum \tilde{p}_i
f(\tilde{\pi}_i)$ is called \textit{optimal}. The minimal length of an
optimal decomposition (the minimal number of pure states in the
decomposition) is $\leq (\mathrm{rank} \rho)^2$ (this is a consequence
of Carath\'eodory's theorem, see Ref.~\cite{Uhlmann1998}, Lemma 1). Since $f
(\rho) $ is an entanglement monotone on $\Omega_\mathcal{H}$ if $f$ is an
entanglement monotone on $\mathcal{H}$
\cite{Vedraletal1997,Vidal2000}, the convex-roof extension is the
standard way of applying a pure-state entanglement measure to mixed
states. For further properties of the convex-roof extension, see
\cite{Uhlmann1998,Uhlmann2000,Vidal2000}. However, as
mentioned, finding the minimum in
Eq.~\eqref{eq:definition_convex_roof} is difficult in general.

\subsection{The convex-roof extension of homogeneous invariants}

Let us now investigate how the convex-roof extension of $\tau$ transforms under the action of $A$, that is, 
how $\tau (\rho)$ and $\tau(\rho^\prime)$ are 
related if $\rho^\prime = L \rho L^\dagger /\mathrm{tr} L\rho L^\dagger$ $(\rho, \rho^\prime \in \Omega_\mathcal{H})$. Interestingly, it turns out that the convex-roof extension 
does not lead to a simple generalization of Eq.~\eqref{eq:result_pure_states} for all $\eta$. Rather, the convex-roof extension singles out invariants of homogeneous degree $\eta = 2$. Only in that case the convex-roof extension of $\tau$ generally can be rescaled as for pure states, 
\begin{align}\label{eq:result}
\tau(\rho^\prime) = \dfrac{\tau (\rho)}{\mathrm{tr} L \rho L^\dagger}\ \ .
\end{align}
This formula represents the case $\eta =2$ of a naive generalization of
Eq.~\eqref{eq:result_pure_states} to $\tau(\rho^\prime) \stackrel{?}{=} \tau
(\rho)/(\mathrm{tr} L \rho L^\dagger)^{\eta/2}$ which, however, is not correct in
general. The main conclusions of this article are based on formula
\eqref{eq:result}. We will see that it can be understood as a
consequence of the fact that $L$ maps an optimal decomposition of $\rho$
onto an optimal decomposition of $\rho^\prime$ if $\eta=2$. For $\eta\neq 2$, the
invariance of $\tau$ under $L$ still guarantees that $\tau(\rho)$ and $\tau(\rho^\prime)$
are either both zero or both nonzero if $\tau(\psi) \geq 0$. These results give
information about the convex roof of $\tau$ for $\rho^\prime$ if it is known for
$\rho$. Applying them to the polynomial invariants allows us to extend
(previous) results for the entanglement of a mixed state as measured
by the polynomial invariants to all states that can be obtained from
this state via general invertible local operations $F$ (ILOs): If $F =
F_1 \otimes \ldots \otimes F_N \in \mathrm{GL}(d_k,\mathbb{C})^{\otimes N}$, then
\begin{align}
\rho^\prime &= \dfrac{F \rho F^\dagger}{\mathrm{tr} F \rho F^\dagger} = \dfrac{A \rho A^\dagger}{\mathrm{tr} A \rho A^\dagger}, 
\end{align}
where $A \in \mathrm{SL}(d_k,\mathbb{C})^{\otimes N}$ is defined as in Eq.~\eqref{genlocop}. 
An example that illustrates how Eq.\ \eqref{eq:result} can be 
utilized to calculate the three-tangle of mixed states will be presented in the next section.

In order to see how Eq.\ \eqref{eq:result} is obtained,
suppose that $\{p_i, \pi_i \}$ and $\{q_j, \varpi_j\}$ are optimal decompositions of $\rho$ and $\rho^\prime$ of lengths $m$ and $n$, respectively. In other words, $\tau(\rho) = \sum_{i=1}^m p_i \tau(\pi_i)$ and $\tau(\rho^\prime) = \sum_{j=1}^n q_j \tau(\varpi_j)$. We express $\rho^\prime$ in terms of the $\{ p_i, \pi_i \}$ and $\rho$ in terms of the $\{ q_j, \varpi_j \}$,
\begin{align}
\rho &= \sum_{i=1}^m{p_i \pi_i} = \sum_{j=1}^n{q_j^\prime \varpi_j^\prime},\label{eq:decs_rho}\\
\rho^\prime &= \sum_{j=1}^n{q_j \varpi_j} =\sum_{i=1}^m {p_i^\prime \pi_i^\prime}, \label{eq:decs_rhoprime}
\end{align}
where
\begin{align}
p_i^\prime &= p_i T^{-1}\mathrm{tr} L \pi_i L^\dagger, \;  & q_j^\prime &= q_j T \mathrm{tr}[ L^{-1} \varpi_j (L^{-1})^\dagger ], \label{eq:transformI} \\ 
\pi_i^\prime &= \dfrac{L \pi_i L^\dagger}{\mathrm{tr} L \pi_i L^\dagger},  \; & \varpi_j^\prime &= \dfrac{L^{-1}  \varpi_j (L^{-1})^\dagger}{\mathrm{tr}[L^{-1}  \varpi_j (L^{-1})^\dagger]}.\label{eq:transformII}
\end{align}
We have used the abbreviation $T= \mathrm{tr} L \rho L^\dagger$. It is easy to see that $\sum_i p_i^\prime = \sum_j q_j^\prime =1$ and that $\pi_i^\prime$ and $\varpi_j^\prime$ are pure states. Since $L \pi_i L^\dagger (=L \pi_i \pi_i^\dagger L^\dagger)$ is a positive operator and $L$ is invertible, $\mathrm{tr} L \pi_i L^\dagger > 0$. Consequently, $T>0$. A similar argument ensures that $\mathrm{tr}[L^{-1}  \varpi_j (L^{-1})^\dagger]>0$. Starting from Eqs.\ \eqref{eq:decs_rho} and \eqref{eq:decs_rhoprime}, one can estimate
\begin{align}
\tau(\rho) &\leq T \sum_{j=1}^n q_j [\mathrm{tr} L^{-1}  \varpi_j (L^{-1})^\dagger]^{\frac{2-\eta}{2}} \tau(\varpi_j), \label{eq:inequalityI}\\
\tau(\rho^\prime) &\leq T^{-1} \sum_{i=1}^m p_i [\mathrm{tr} L \pi_i L^\dagger]^{\frac{2-\eta}{2}} \tau(\pi_i).\label{eq:inequalityII}
\end{align}
If $\eta=2$, Eq.\ \eqref{eq:result} follows. Moreover, one sees
  that $\{p_i^\prime, \pi_i^\prime \}$ is an optimal decomposition of $\rho^\prime$ because
  $\sum_i p_i^\prime \tau(\pi_i^\prime)= \sum_i p_i \tau(\pi_i)/T = \tau(\rho)/T$. Further, $\{q_j^\prime,
\varpi_j^\prime\}$ is an optimal decomposition of $\rho$. Hence, an optimal
decomposition $\{ p_i, \pi_i \}$ ($\{q_j,\varpi_j\}$) of $\rho$ ($\rho^\prime$) is mapped by
$L$ ($L^{-1}$) onto an optimal decomposition of $\rho^\prime$ ($\rho$) which is
given by $\{p_i^\prime, \pi_i^\prime\}$ ($\{q_j^\prime, \varpi_j^\prime\}$) in
Eqs.~\eqref{eq:transformI} and \eqref{eq:transformII}. 

If $\eta \neq 2$, but
$\tau \geq 0$, one can still conclude from Eqs.\ \eqref{eq:inequalityI} and
\eqref{eq:inequalityII} that either both $\tau(\rho)$ and $\tau(\rho^\prime)$ are zero
or both are nonzero. The assumption $\tau(\rho) =0$ implies $\tau(\pi_i) =0$ for
all $\pi_i$ and, due to Eq.\ \eqref{eq:inequalityII}, $\tau(\rho^\prime) =0$. On
the other hand, the assumption $\tau(\rho)>0$ implies that $\tau(\varpi_j)>0$ for at
least one $\varpi_j$ because of Eq.\ \eqref{eq:inequalityI}. The converse
follows analogously.

Let us briefly discuss these results with regard to the polynomial invariants and 
their application as entanglement measures. As already indicated, our results, in 
particular Eq.\ \eqref{eq:result}, provide a useful tool for learning about the 
convex-roof extension of a polynomial invariant for a mixed state $\rho^\prime$ 
if it is already known for a state $\rho$ that can be transformed into $\rho^\prime$ 
via SLOCC. However, we have also gained another important insight about the convex-roof 
extension of polynomial invariants: It has peculiar properties for polynomial entanglement 
measures of homogeneity degree 2 (or, of linear degree if the pure-state measures 
are regarded as functions of rank-one density operators) and therefore indicates a 
special significance of the latter. Only for entanglement measures of this 
type, the invariance property for normalized pure-states generalizes to normalized 
mixed states and the entanglement both of pure and mixed SLOCC-equivalent states is 
related by a simple analytical formula, Eq.~\eqref{eq:result}.


\section{Examples}

Now we consider several examples to illustrate how the rescaling
method from Section \ref{sect:rescalingmethod} can be applied to generalize existing
results for mixed-state entanglement. We will also demonstrate what
happens if one tries to apply it to monotones of degree $\neq 2$.

The examples we consider are mixed states of three qubits. For
three-qubit systems, there is only one polynomial
$\mathrm{SL}(2,\mathbb{C})^{\otimes3}$ invariant, the three-tangle
\cite{CKW2000}. For $\ket{\psi}=\sum_{i,j,k=0}^1\psi_{ijk}\ket{ijk}$ it
is given by
\begin{eqnarray}
  \label{eq:three-tangle}
  \tau_3(\psi) &=& 4\left|d_1 - 2 d_2 + 4 d_3\right|,\nonumber\\
  d_1 &=& \psi_{000}^2\psi_{111}^2 + \psi_{001}^2\psi_{110}^2 + \psi_{010}^2\psi_{101}^2
  + \psi_{011}^2\psi_{100}^2\nonumber\\
  d_2 &=& \psi_{000}\psi_{001}\psi_{110}\psi_{111} + \psi_{000}\psi_{010}\psi_{101}\psi_{111} +
 \nonumber\\ &&
  + \psi_{000}\psi_{011}\psi_{100}\psi_{111}
  + \psi_{001}\psi_{010}\psi_{101}\psi_{110} + 
 \nonumber\\ &&
  + \psi_{001}\psi_{011}\psi_{100}\psi_{110} +
  \psi_{010}\psi_{011}\psi_{100}\psi_{101}\nonumber\\
  d_3 &=& \psi_{000}\psi_{110}\psi_{101}\psi_{011} + \psi_{100}\psi_{010}\psi_{001}\psi_{111}
\end{eqnarray}
and distinguishes Green\-ber\-ger-Horne-Zei\-lin\-ger (GHZ) 
entanglement from $W$-type entanglement. For
mixed states, it is defined by convex-roof extension.

The three-tangle has homogeneous degree $4$. However to apply the
rescaling method Eq. (\ref{eq:result}) as described above, 
degree $2$ is required. Therefore on pure states we have to use the square root of
the three-tangle, $\tau(\psi)=\sqrt{\tau_3(\psi)}$ which is an
entanglement monotone as well~\cite{Duer2000}. For mixed states, we again define
$\tau(\rho)$ through the convex-roof extension. Note that this is \emph{not}
the same as taking the square root of $\tau_3(\rho)$. However $\tau(\rho)=0$ iff
$\tau_3(\rho)=0$; therefore both are equally suitable to distinguish
GHZ-type and $W$-type entanglement.

\subsection{Mixtures of generalized GHZ and $W$ states}
\label{sect:mixtures}

We consider $\tau(\rho)$ for rank-2 mixtures
\begin{equation}
\label{eq:gmix}
\rho(p) = p\ket{\mathrm{gGHZ}}\!\bra{\mathrm{gGHZ}}
+ (1-p)\ket{\mathrm{g}W}\!\bra{\mathrm{g}W}
\end{equation}
of a generalized GHZ and generalized $W$ state (both normalized)
\begin{eqnarray}
  \label{eq:gGHZ}
  \ket{\mathrm{gGHZ}}\ &=&\ a\ket{000} + b\ket{111} \nonumber\\,
  \ket{\mathrm{g}W}\ &=&\ c\ket{100} + d\ket{010} + e\ket{001}\ \ .
\end{eqnarray}
The three-tangle of those states has been calculated in \cite{Eltschka2008}.
We will be able to reuse pure-state results from that paper.

We begin by solving the problem using the characteristic curve method
\cite{OSU2008} and then demonstrate how those solutions can be mapped
to each other by means of the rescaling method.

The characteristic curve is defined as the minimum tangle of the
states at the same ``height'' in the Bloch sphere as the corresponding
mixed state,
\begin{equation}
  \label{eq:charcurve}
  \tilde \tau(p) = \min_\phi(\tau(\sqrt{p}\ket{\mathrm{gGHZ}} +
  \sqrt{1-p}\rme^{\rmi\phi}\ket{\mathrm{g}W})).
\end{equation}
Since we are dealing with pure states at this point, and the square
root is monotonic, we can reuse the results of Ref.~\cite{Eltschka2008} and get
\begin{equation}
  \label{eq:roottau3}
  \tilde \tau(p) = 2\left|ab\right|\sqrt{\left|p^2-|s|\sqrt{p(1-p)^3}\right|}
\end{equation}
where we have used the definition
\begin{equation}
  \label{eq:s}
  s = \frac{4cdf}{a^2b} \ \ .
\end{equation}

The \emph{convex} characteristic curve $\tau_s^*(p)$ is the function
convex hull of $\tilde \tau(p)$, i.\ e., the largest convex function which is
nowhere larger than $\tilde \tau(p)$. As shown in Ref.~\cite{OSU2008}, the
convex characteristic curve is always a lower limit to $\tau(p)$. Using
the results in \cite{Eltschka2008}, we find that $\tilde \tau(p)$ has zeros at
$p=0$ and at $p=p_0$ with
\begin{equation}
\label{eq:p0}
p_0=\frac{|s|^{\frac{2}{3}}}{1+|s|^{\frac{2}{3}}}\ \ .
\end{equation}
It is easy to check that for $p>p_0$, $\tilde \tau(p)$ is concave (note
that this is different from the three-tangle itself where $\tilde
\tau_3(p)$ is convex right above $p_0$ for $s<2\sqrt{2}$). Therefore, in
this region, the convex characteristic curve is a straight line. 
We thus get
\begin{equation}
  \label{eq:convcharcurve}
  \tau^*(p) =
  \begin{cases}
    0 & \mbox{for } 0 \leq p \leq p_0,\\
    2\left|ab\right|\left(p-(1-p)|s|^{\frac{2}{3}}\right) & \mbox{for } p_0 \leq p \leq 1.
  \end{cases}
\end{equation}

In the current case, the convex characteristic curve actually gives
the correct value of $\tau(p)$. This can be seen from the symmetry of the
problem: Applying the local unitary transformation
\begin{equation}
  \label{eq:GHZWsymm}
  U_3 = \exp\left[\frac{\rmi\pi}{3} \sigma_z^{(1)}\right]\exp\left[\frac{\rmi\pi}{3} \sigma_z^{(2)}\right]\exp\left[ \frac{\rmi\pi}{3} \sigma_z^{(3)}\right]
\end{equation}
changes neither the GHZ nor the $W$ state (and thus also not their
mixtures) and, of course, nor the three-tangle. 
But it changes the relative
phase between the GHZ and the $W$ state in the superpositions of Eq.~\eqref{eq:charcurve},
thus rotating the Bloch sphere spanned by these
superpositions by $2\pi/3$. Thus each state gives rise under this
symmetry $U_3$ to three locally SU equivalent states whose equal
mixture gives exactly $\rho (p)$ (Eq.~\eqref{eq:gmix}). This is especially true
for the minimum $\tau$ state defining the characteristic curve. Thus, due
to the symmetry the characteristic curve gives an upper limit to $\tau$.

Since the convex characteristic curve always provides a lower limit,
wherever both agree, the unconvexified curve gives the correct
tangle, and therefore in those points the above decomposition is an
optimal one.

While the above argument applies only to the points $p=0$, $p=p_0$, and
$p=1$, it is obvious that for the linear sections of the convex
characteristic curves, the value $\tau^*(p)$ can be achieved by
appropriately combining the optimal decompositions of the
corresponding end points which proves that indeed $\tau(\rho(p))=\tau^*(p)$.

After deriving the solution of the problem, we now show that different states
\eqref{eq:gmix} are SLOCC equivalent and, hence, that the solutions 
$\tau$ for different values of $a,\ldots,f$ can be obtained from each other via
the rescaling method. To this end,
we apply a diagonal invertible local operation of determinant one,
\begin{equation}
  \label{eq:gILO}
  A = \diagSL{\alpha}\otimes\diagSL{\beta}\otimes\diagSL{\gamma},
\end{equation}
which transforms \ket{\mathrm{gGHZ}} to $\mathcal{N}_1\ket{\mathrm{GHZ}}$ and
\ket{\mathrm{g}W} to $\mathcal{N}_2\ket{W}$. An easy calculation shows
that this is achieved for
\begin{equation}
  \label{eq:gILOparms}
  \alpha = \left(\frac{b c^2}{a d f}\right)^{\frac{1}{6}},\quad
  \beta = \left(\frac{b d^2}{a c f}\right)^{\frac{1}{6}},\quad
  \gamma = \left(\frac{b f^2}{a c d}\right)^{\frac{1}{6}}.
\end{equation}
This transforms \ket{\mathrm{gGHZ}} into
$\sqrt{2ab}\ket{\mathrm{GHZ}}$ and \ket{\mathrm{g}W} into
$\sqrt{3}\left(\sqrt{\frac{b}{a}}cdf\right)^{\frac{1}{3}}\ket{W}
=\sqrt{\frac{3}{8}}\cdot\sqrt{2ab}\cdot(2s)^{\frac{1}{3}}\ket{W}$
and therefore $\rho$ into
\begin{equation}
  \label{eq:rhotildeunnormed}
  A\rho A^\dagger =
  2\left|ab\right|\left(p\ket{\mathrm{GHZ}}\!\bra{\mathrm{GHZ}}
  + \frac{3(1-p)}{8}\left|2s\right|^{\frac{2}{3}}\ket{W}\!\bra{W}\right).
\end{equation}
Since the trace of this operator is
\begin{equation}
\label{eq:T}
T\ \equiv\ \tr(A\rho A^\dagger) = 2\left|ab\right|\left(p +
\frac{3(1-p)}{8}\left|2s\right|^{2/3}\right),
\end{equation}
we have to divide $\rho'$ by this expression and find
\begin{equation}
  \label{eq:rhotilde}
  \rho' = \frac{A\rho A^\dagger}{T}
  = p' \ket{\mathrm{GHZ}}\!\bra{\mathrm{GHZ}}
  + (1-p') \ket{W}\!\bra{W}
\end{equation}
with
\begin{equation}
  \label{eq:ptilde}
  p^{\prime}\ = \ \frac{p}{p+\frac{3(1-p)}{8}|2s|^{2/3}}
\end{equation}
or, solving for $p$,
\begin{equation}
  \label{eq:pfromptilde}
  p\ =\ \frac{3p' |2s|^{2/3}}{3 p' |2s|^{2/3}+8(1-p')}
\end{equation}
Note that this is an increasing function of $p'$.

Now we can apply the rescaling method described in the previous section to this
problem. By using Eq. (\ref{eq:result}) we obtain
\begin{equation}
  \label{eq:roottrans}
  \tau(\rho(p))
 = T\ \tau(\rho'(p')),
\end{equation}
where $T$ is taken from Eq. (\ref{eq:T}), and $p'$ from
(\ref{eq:ptilde}).

\begin{figure}[ht]
  \centering
  \includegraphics[height=0.265\textwidth]{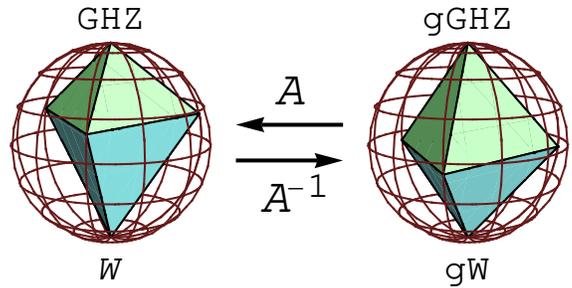}
  \caption{ Illustration of the rescaling method for the square root
    of the three-tangle $\tau$ of mixtures of standard (left) and
    generalized (right) GHZ and $W$ states. The two orthogonal
    eigenstates (g)GHZ and (g)$W$ are located at the north and south
    poles of the Bloch spheres, respectively, and the mixing
    parameters $p^\prime$, $p$ vary along the lines connecting the two
    poles. The symmetric GHZ/$W$ problem (left) can be transformed into
    the generalized GHZ/$W$ problem (right) and vice versa by an
    invertible local operation. The corners of the tetrahedra show the
    pure states of the optimal decomposition of the states inside
    those tetrahedra. In the lower (blue) tetrahedron, $\tau=0$, while in
    the upper (green) tetrahedron, it grows linearly with $p^\prime$ ($p$).
    If the convex-roof extension is known for one of the problems, it
    can be calculated for the other by virtue of the rescaling
    method.}
  \label{fig:degree2}
\end{figure}

For the standard GHZ/$W$ problem, that is, $a=b$ and $c=d=f$, we have
$s=2^{7/2}/3^{3/2}$ and $2|ab| = 1$, and therefore,
\begin{equation}
  \label{eq:taustandardGHZW}
  \tau(\rho'(p')) =
  \begin{cases}
    0 & \mbox{for } 0 \leq p' \leq p'_0 =
    \frac{2^{7/3}}{3+2^{7/3}},\\
    p'-(1-p')\frac{2^{7/3}}{3} & \mbox{for } p'_0 \leq p' \leq 1.
  \end{cases}
\end{equation}
By inserting all values one indeed recovers Eq.~(\ref{eq:convcharcurve})
with the correct value of $p_0$, Eq. (\ref{eq:p0}), see Fig.\ \ref{fig:degree2}. This means in turn
that, by virtue of the rescaling method, the knowledge of the solution of the standard problem 
with symmetric
GHZ and $W$ states in Eq.~\eqref{eq:taustandardGHZW} is sufficient to obtain 
the solution of the more general states in Eq.~\eqref{eq:gmix}. This is achieved 
simply by applying ILOs according to Eq.~\eqref{eq:gILO} and appropriate rescaling following
Eq.~\eqref{eq:result}.

Note that while we have considered only diagonal transformations, 
the method is of course valid for general invertible local operations. 
Those break the
symmetry noted above and therefore lead to cases which cannot be
solved using the convex characteristic curve method. For those
problems, no general method is yet known.

\subsection{Using the rescaling method for degrees other than $2$}

As mentioned before, the rescaling method is not fully
applicable to entanglement measures with homogeneous degree $\eta\neq 2$. 
However, by considering the three-tangle ($\eta=4$)
of GHZ/$W$ mixtures we will see that
if the problem has certain symmetries, the rescaling method
can still be useful for calculating entanglement measures
with $\eta \neq 2$.

\begin{figure}[hb]
  \centering 
  \includegraphics[trim = 1cm 0cm 1cm 0cm, height=0.18\textwidth]{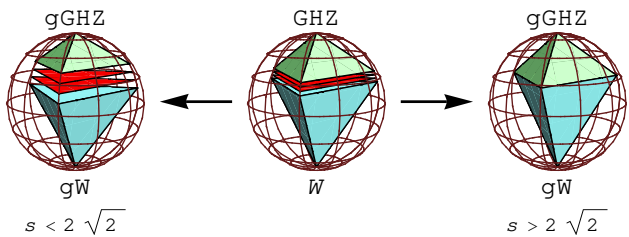}
  \caption{Illustration of the three-tangle $\tau_3$ of mixtures of
    standard GHZ and $W$ states and generalized GHZ and $W$ states. As
    in Fig.~\ref{fig:degree2}, the (g)GHZ and (g)$W$ states are at the
    north and south poles of the Bloch spheres, respectively, and the
    corners of the tetrahedra and triangles give the elements of the
    optimal decompositions. In each red triangle,
    $\tau_3=\mathrm{const}$. In the blue tetrahedra, $\tau_3=0$, and in the
    green tetrahedra $\tau_3$ depends linearly on the mixing parameter.
    Here, the sphere in the center represents the standard GHZ/$W$
    case. The transformation to the generalized problem leads to two
    different cases: For $s<2\sqrt{2}$ (left sphere), the structure of
    the symmetric case is preserved. However, for $s>2\sqrt{2}$, the
    two tetrahedra sit directly on top of each other. As  states
      equivalent under ILOs can have optimal decompositions for $\tau_3$
    with different lengths, the figure indicates that the rescaling
    method cannot be applied to calculate $\tau_3$.}
  \label{fig:degree4}
\end{figure}

The three-tangle of the mixtures in Eq.~(\ref{eq:gmix}) has already
been calculated in \cite{Eltschka2008} and, for the special case of mixing
standard GHZ and standard $W$, in \cite{Lohmayer2006} 
(see also Fig.\ \ref{fig:degree4}).
For the mixtures of standard GHZ and standard $W$, the three-tangle has
the form
\begin{equation}
  \label{eq:3tangleGHZW}
  \tau_3(p')=
  \begin{cases}
    0, &  0 \leq p' \leq p'_0,\\
    p'^2-\frac{2^{7/2}}{3^{3/2}}\sqrt{p'(1-p')^3}, &
    p'_0 \leq p' \leq p'_1,\\
    1-(1-p')\left(\frac{3}{2}+\frac{1}{18}\sqrt{465}\right), &
    p'_1 \leq p' \leq 1,
  \end{cases}
\end{equation}
where $p'_0=\frac{2^{7/3}}{3+2^{7/3}}$ and
$p'_1=\frac{1}{2}+\frac{1}{18}\sqrt{465}$. We now derive the results
of a simple-minded application of the rescaling method to the three-tangle
which is based on the naive generalization of Eq.\
\eqref{eq:result_pure_states} (below Eq.\ \eqref{eq:result}) with $\eta =
4$. That is, we calculate $p'$ from $p$ using Eq.~(\ref{eq:ptilde}),
apply the function above, and then multiply with $T^2$ from Eq.
(\ref{eq:T}). It will turn out that this procedure indeed leads to
incorrect results in general.This is also suggested by
Fig.~\ref{fig:degree4}: Optimal decompositions of states which
  can be transformed into one another by means of operations
  Eq.~(\ref{eq:gILO}) can have different lengths. Hence, the ILOs in
  general do not map optimal decompositions onto optimal
  decompositions, and one would therefore not expect that the
three-tangle of a mixed state can be traced back to the (known)
three-tangle of an SLOCC-equivalent state via the rescaling method.
However, we will also see that in some cases the correct three-tangle
of mixtures of generalized GHZ and $W$ states can be obtained from
Eqs.\ \eqref{eq:3tangleGHZW} by means of this method due to the
particular symmetries of this problem.

To identify the range of $p$ where the three-tangle vanishes, we
insert $p'_0$ into Eq. (\ref{eq:pfromptilde}) and get 
Eq.~(\ref{eq:p0}) in agreement with Ref.~\cite{Eltschka2008}. Indeed this was to be
expected, because the method allows us to reliably distinguish between
zero and nonzero values of the polynomial even for homogeneity
degrees other than 2.

Next, we calculate the tangle for $p_0\leq p\leq p_1$ using the method above, where 
$p_1$ is the $p$ value beyond that $\tau_3(\rho(p))$ increases 
linearly~\cite{Eltschka2008}. From
\begin{equation}
  \label{eq:tau3rho}
  \tau_3(\rho(p)) = T^2 \tau_3(\rho'(p'))
\end{equation}
we get
\begin{equation}
  \label{eq:taucurve}
  \tau_3(p) = 4\left|ab\right|^2\left(p^2 - |s|\sqrt{p(1-p)^3}\right)
\end{equation}
which again agrees exactly with the result in Ref.~\cite{Eltschka2008}. This comes
somewhat as a surprise, because the method should not be applicable
here. However, it is easily explained by considering the symmetry of
the problem.

Let us consider the special case that both $\{p_i,\pi_i\}$ and $\{p_i, A \pi_i
A^\dagger/T\}$ are optimal decompositions of $\rho$ and $\rho^\prime$ in normalized pure
states. This implies $\mathrm{tr}( A \pi_i A^\dagger) = T$ for all $i$. One
easily sees from Eqs.\ \eqref{eq:inequalityI} and
\eqref{eq:inequalityII} that, under these conditions, $\tau(\rho) = T^{\eta/2}
\tau(\rho^\prime)$ if $\tau$ is an arbitrary homogeneous invariant, or, in the case of
the three-tangle, $\tau_3(\rho) = T^{2} \tau_3(\rho^\prime)$. The specific mixed states
and ILOs of our example satisfy these conditions for $p_0\leq p\leq p_1$, as
we now explain.

In this region characteristic curve and convex characteristic curve
agree with each other. As described above, this is because there is an
optimal decomposition which consists purely of states related by the
local symmetry (\ref{eq:GHZWsymm}) which rotates the Bloch sphere by
$2\pi/3$.

Arbitrary rotations of the Bloch sphere about the GHZ/$W$ axis cannot
be achieved by local operations; however, they can be realized by
nonlocal diagonal unitary transformations, e.g.,
\hfill $U(\varphi)=\diag(1,\rme^{\rmi\varphi},\rme^{\rmi\varphi},\rme^{-\rmi\varphi},$
$\rme^{\rmi\varphi},\rme^{-\rmi\varphi},\rme^{-\rmi\varphi},1)$.
Note that the local operations (\ref{eq:GHZWsymm}) are a subset of
these. Now the ILOs (\ref{eq:gILO}) are also diagonal and therefore
commute with those transformations. This means especially that if two
vectors are equivalent under this symmetry $U(\varphi)$, the transformed vectors
are also equivalent under this symmetry, which means that horizontal
planes in the Bloch sphere are just moved vertically by the transformation.
Moreover, it is easy to verify that if two pure states can be
transformed into each other by $U(\varphi)$, i.e., share the same latitude,
their norms are also multiplied by the same factor. This implies that
\emph{for those states}, the transformed state has minimal three-tangle
iff the original state had minimal three-tangle.
Of course, all convex combinations of those states are also multiplied
with the same factor. Especially, the characteristic curve of the
original problem is mapped onto the characteristic curve of the
transformed problem. 

However, to get the correct three-tangle, one needs the \emph{convex}
characteristic curve, which does not agree with the characteristic
curve above $p_1$. Now we cannot expect that the transform of the
latter gives the same result as the convexification of the transformed
curve and that optimal decompositions are mapped to optimal decompositions in general. Indeed, trying a straightforward calculation of $p_1$
using Eq.~(\ref{eq:pfromptilde}), we obtain
\begin{equation}
  \label{eq:p1}
  p_1^{\text{Eq. (\ref{eq:pfromptilde})}} 
       = \frac{24\ |2s|^{\frac{2}{3}}}{91-3 \sqrt{465}+24\ |2s|^{\frac{2}{3}}}
\end{equation}
which not only does not agree with the correct $p_1$ from \cite{Eltschka2008},
but even has a wrong behavior. While the correct $p_1$ decreases
from $1$ to $1/2$ on increasing $s$ (actually, only until it hits
$p_0$), $p_1^{\text{Eq. (\ref{eq:pfromptilde})}}$  instead increases
from $0$ to $1$. 

As the unconvexified characteristic curve is mapped
onto the unconvexified characteristic curve of the transformed problem, 
the correct three-tangle can be obtained by using the former to
calculate the convex characteristic curve directly.
However, we emphasize again that this is only due to the
high symmetry of the problem and the fact that the symmetry operations 
commute with the diagonal ILOs.

As in the case of degree 2 (Section \ref{sect:mixtures}), one might consider to further 
extend the range
of solutions derived from the standard GHZ/$W$ case by applying
arbitrary (i.e., not only diagonal ones) local SL transformations. 
However, general
transformations modify the unitary symmetry of the problem into 
one which is nonunitary, $AU(\varphi)A^{-1}$. 
Such a transformation in general does not map the characteristic curve (and
the optimal decompositions) of the
standard GHZ/$W$ mixtures onto the characteristic curve (and the optimal
decompositions) of the transformed mixed states. Therefore, in general the rescaling method
is not useful for finding new solutions for the convex roofs of entanglement
measures with homogeneity degree $\eta \neq 2$.   
However, note that the zero polytope of the original problem
will be mapped exactly onto the zero polytope of the transformed problem,
irrespective of $\eta$ (see Section \ref{sec:rescalingmethod}).

\subsection{Using the rescaling method for mixed states of rank greater than $2$}

If the degree of homogeneity of the polynomial invariant is 2, the rescaling method
is fully applicable. We show now by discussing further examples that it can be also successfully 
employed for calculating the entanglement of mixed states with rank $>2$.
Consider, for instance, 
rank-3 mixtures of \ket{\mathrm{GHZ}}, \ket{W} and $\ket{\bar
  W}=(\ket{011}+\ket{101}+\ket{110})/\sqrt{3}$. The three-tangle for
such mixtures has been found by Jung {\em et al.}~\cite{Jung2009}. 
We illustrate the application of 
the rescaling method for states of rank $>2$ by briefly sketching
how the square root of the three-tangle can be obtained for
those states and all SLOCC equivalent states.
Adopting the
parameterization from Ref.~\cite{Jung2009}, we have
\begin{equation}
  \label{eq:rhoGHZWWquer}
  \rho(p,q) = p\ket{\mathrm{GHZ}}\!\bra{\mathrm{GHZ}} + q\ket{W}\!\bra{W} +
  (1-p-q)\ket{\bar W}\!\bra{\bar W}.
\end{equation}
The analog of the characteristic curve is now the two-dimensional
``characteristic surface''
\begin{eqnarray}
  \tilde \tau(p,q) \ & =& \ \min_{\phi_1,\phi_2}
  \tau\left[\sqrt{p}\ket{\mathrm{GHZ}} - \right.\nonumber\\
                   && \left. 
                     - \sqrt{q}\rme^{\rmi \phi_1}\ket{W} -
  \sqrt{1-p-q}\rme^{\rmi \phi_2}\ket{\bar W}\right],
  \label{eq:charsurf-GHZWWquer}
\end{eqnarray}
where $\tau=\sqrt{\tau_3}$. We have again adopted the parameterization of the states used in
\cite{Jung2009}. As before, the problem has the symmetry
(\ref{eq:GHZWsymm}), and an equal mixture of a set of states
equivalent to a given state under this symmetry states gives $\rho(p,q)$.
It is easy to see that $\tau$ takes its minimal value
for $\phi_1=\phi_2=0$, which results in
\begin{equation}
  \label{eq:taupq}
  \tilde \tau(p,q) = \sqrt{\left|p^2 - 4p\sqrt{qr} -
      \frac{4}{3}qr -
      \frac{8\sqrt{6}}{9}\left(\sqrt{pq^3}+\sqrt{pr^3}\right)\right|},
\end{equation}
where we have defined $r=1-p-q$. The decompositions of $\rho(p,q)$ leading to those values (note that 
in general they do not represent the optimal decompositions because the 
surface is not convex) are the ones given in Eq.~(11) of \cite{Jung2009}.

\begin{figure}[b]
  \centering
  \includegraphics[width=0.3\textwidth]{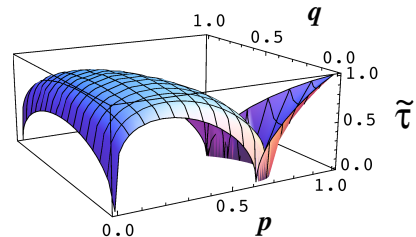}
  \caption{ Characteristic surface $\tilde\tau(p,q)$ (Eq. (36))
    for mixtures of standard GHZ, standard $W$ and flipped $W$ state.
    As described in the text, due to the symmetry of the problem, the
    convex hull of this function gives the correct value of 
    $\tau(\rho)$. The zeros of
    $\tau$ are found for the $W$ state ($p=0$, $q=1$), the flipped $W$
    state ($p=0$, $q=0$), and for a continuous curve starting at
    $p=q=p'_0$ (see Eq.~(27)) and ending at $p=p'_0$, $q=0$, where the
    characteristic surface has a kink. The convex hull of this
    function is zero between the line connecting $W$/flipped $W$ and
    the curved line of zeros described before. 
    From the graph it is also easy to see that between that
    curve and the GHZ state, the convex hull consists of the straight
    line segments which connect the individual points of the curve with
    the point corresponding to the GHZ state.}
  \label{fig:charsurf}
\end{figure}

When plotting this function (see Fig.~\ref{fig:charsurf}), one easily
sees that there is a curved line of zeros. All states between that
line and the connecting line between \ket{W} and \ket{\bar W} are $W$-type
states with $\tau=\tau_3=0$. 
One also sees in the graph in Fig.~\ref{fig:charsurf} that
beyond this line  of zeros, one can convexify the surface by just connecting the
points of the zero line with the GHZ state using straight lines.
Calculating the exact values lies beyond the scope of the present paper
as our aim merely is to illustrate the potential of the application 
of the rescaling method.
We note, however, that several points on the line of zeros are given in
Table I of Ref.~\cite{Jung2009}.
Once the solution for the convex roof of $\tau=\sqrt{\tau_3}$ is obtained, 
it is straightforward to apply the criterion (\ref{eq:result}).
This allows us to calculate the square root of the
three-tangle for a larger set of states.

We conclude this section by noting that  also in the case of these rank-3 mixtures
even the three-tangle $\tau_3$ (degree 4) could be extended by a modified application of rescaling.
This is (in complete analogy with the previous section) because the states in 
\eqref{eq:charsurf-GHZWWquer} whose $\tau_3$ is minimized 
for the characteristic surface are equivalent under a diagonal (nonlocal)
unitary symmetry. 
Therefore, by considering only diagonal ILOs, the
characteristic surface of the three-tangle (without square root)
is again mapped to the characteristic surface of the transformed
problem. Thus, in analogy to the preceding section,
we can calculate the three-tangle ``after ILOs'' also for
this case from the three-tangle of the original problem, by
convexifying the transformed surface.

Finally, we remark that there are cases of even higher rank where 
the entanglement of classes of SLOCC-equi\-valent states can be obtained
via the rescaling method. 
The three-tangle of certain mixtures of three-qubit GHZ states of high
rank has been calculated by Shu-Juan et al.\ \cite{Fei2011}.
It should be straightforward to carry out also those calculations for the
square root of the three-tangle. Then the rescaling method can be used
to extend the scope of those results. Since Shu-Juan et al.
used the convex characteristic curve method, 
it should also be possible to
calculate, for a suitable restriction of the local transformations, the characteristic 
surface of the transformed problem for
the three-tangle. Its convexification should yield the three-tangle of the
transformed problem.

\section{Conclusion}

We have developed a rescaling method to calculate poly\-nomial-based
entanglement monotones of homogeneous degree 2 (in the coefficients
of a pure quantum state) in mixed states. 
The method is based on transforming the mixed state under consideration 
into one of an already solved problem by using invertible local operations.
We have further demonstrated that rescaling is not generally applicable to functions of
homogeneous degrees other than 2. Therefore, from the point of view of 
the convex-roof
construction, there is a clear preference for 
entanglement monotones of homogeneous degree 2 (and correspondingly,
of homogeneous degree 1 in the coefficients of the density matrix). 
This is very much in line
with some conclusions by Verstraete et al.~\cite{Verstraete2003}.

The application of the rescaling method has been illustrated
by means of several examples. We have found that the presence of a 
certain symmetry in some cases even allows us to calculate the convex roof 
of entanglement measures with homogeneous degrees
other than two. Moreover, the question whether the monotone is zero or
nonzero, which is important for SLOCC classification, can be reliably
answered with this method independent of the homogeneous degree.

\vspace{1mm}
{\em Note added:} After completion of this work we became aware
of Ref.\ \cite{Gour2010}  that also contains a proof of Eq.\ \eqref{eq:result}
and applications thereof.

\vspace{1mm}
{\em Acknowledgments. --- }
This work was supported by the German Research Foundation within SFB 631 
and SPP 1386 (CE) and Basque Government grant IT-472 (JS).
The authors thank A.\ Uhlmann for helpful comments and 
J.\ von Delft, J.\ Fabian, F.\ Marquardt, 
and K.\ Richter for their support of this research.

\end{document}